\documentstyle[aps,prl,multicol,psfig]{revtex}

\newcommand{\dt }{\mbox{d}t}
\newcommand{\bra}[1]{\left<#1\right|} 
\newcommand{\ket}[1]{\left|#1\right>} 
 
\newcommand{\ketbra}[2]{\left|#1\right>\!\!\left<#2\right|} 
\newcommand{\ketbrad}[1]{\left|#1\right>\!\!\left<#1\right|} 
\newcommand{\Tr}[1]{\mbox{Tr}\left( #1 \right)} 
\newcommand{\eqref}[1]{(\ref{#1})}

\begin{document}
\title{Removal of a single photon by adaptive absorption}
\author{John Calsamiglia$^1$, Stephen M.~Barnett$^2$, 
Norbert L\"utkenhaus$^3$, Kalle-Antti Suominen$^{1,4}$}
\address{$^1$  Helsinki Institute of Physics, PL 64, 
FIN-00014 Helsingin yliopisto, Finland\\
$^2$ Department of Physics and Applied Physics,
 University of Strathclyde,John Anderson Building,
107 Rottenrow,Glasgow G4 0NG, Scotland.\\
$^3$ MagiQ Technologies Inc.,
275 Seventh Avenue, 26th floor, NY 10001-67 08, USA.\\
$^4$ Department of Applied Physics, University of Turku,
FIN-20014 Turun yliopisto, Finland}

\date{Received \today}
\maketitle
\begin{abstract}
We present a method to remove, using only linear optics, exactly one 
photon from a field-mode. This is achieved by putting the system in contact 
with an absorbing environment which is under continuous monitoring. 
A feedback mechanism then decouples the system from 
the environment as soon as the first photon is absorbed. 
We propose a possible scheme to implement this process and 
provide the theoretical tools to describe it.
\end{abstract}
\begin{multicols}{2}
%\narrowtext

\section{Introduction}

In the last decades experimental technology has 
reached the point in which quantum effects cannot only be observed but 
also controlled. Sensors have become very precise and 
individual quantum systems can be manipulated  
through  coupling to semiclassical driving fields.  
Despite this, we are still far from having full control over quantum 
systems.  In the field of 
quantum optics, the two main reasons for this are that the coupling to the
driving fields is in practice quite limited, and  the
non-desired coupling to the environment
produces damping and decoherence effects which hinder controllability.
Classical control theory has been quite successful in manipulating
classical systems with complex or partially unknown dynamics. 
Classical control consists of closely monitoring the 
system, processing the measured data and 
applying the appropriate feedback signal to modify the evolution of the 
system. In the field of quantum optics this idea has been used
to stabilize the phase or intensity of lasers. 
However, this reduction of noise  could
be explained classically. 

Non-linear effects are used to 
reduce noise in the quantum domain. 
Yamamoto \emph{et al}\cite{yama86} presented the 
first scheme which relied only on feedback to generate photocurrents 
with noise below the standard quantum limit. Based on Langevin equations,
Shapiro \emph{et al.} \cite{shapiro87} gave the first
quantum and semiclassical descriptions of feedback control.
Wiseman and coworkers \cite{wiseman93} derived a closed-loop 
master equation describing the instantaneous feedback of a 
homodyne current measured outside an optical cavity. 
This work, based on quantum trajectories, made clear the connection 
between control and quantum measurement theory and made possible a 
complete understanding of `quantum-limited' feedback.
Subsequently several applications and elaborations of this first quantum 
theory of continuous feedback appeared. In general the feedback does 
not have to be instantaneous and can be a functional 
of the entire history of measurement 
results, thus typically resulting in a non-Markovian evolution. 
Doherty and Jacobs \cite{doherty99} showed that in 
determining the form of this functional it is useful to formulate 
feedback control in two steps: estimation of the state of the system, 
and application of appropriate control inputs to affect the dynamics.
In \cite{doherty00}  Doherty \emph{et al.} extended  this idea further
to investigate how methods in the well developed classical control theory 
can apply quantum feedback control.  
From the application point of view, feedback has been proposed as a means 
to improve atom localization in standing waves \cite{dunn97,doherty99}, cooling 
of cavity mirrors \cite{mancini97}, controlling coherence of two-level 
atoms \cite{hofman98}, inverting quantum jumps \cite{mabuchi96},
line narrowing in atomic fluorescence \cite{wiseman98},
slowing down of decoherence effects \cite{tombesi95}, 
  and preserving macroscopic superpositions of light fields \cite{light}.

In most of these applications, feedback is used to keep  an open 
system under control. The back-action of measurement in 
the `quantum-limited' feedback  is nearly considered as a drawback, and the 
feedback is enforced by the action of a driving field that shapes the 
potential.
In our proposal the feedback itself is used to create an otherwise 
difficult operation. In fact, the evolution is tailored only by the continuous 
measurement procedure. 

The idea in this work is to show that feedback control 
implemented by simple linear optical elements can be of great use
in manipulating quantum states of light. 
This is, at present, particularly interesting in the context of quantum 
communication where photons are the only serious qubit carriers.
Their very weak coupling strengths suppress decoherence effects,
but at the same time render rather difficult the 
implementation of quantum gates. Any tools which perform non-trivial
tasks without relying on photon-photon interactions are therefore 
very desirable. The use of linear optical elements has 
already proven to facilitate some tasks \cite{lopt}. 
Here, we address the particular problem of removing 
precisely one photon from a field mode. 
Our proposal is based on a very simple form of 
feedback and for its  implementation only linear optical elements are needed.
The central idea is that of \emph{adaptive absorption} and consists of 
coupling the  mode to an environment and monitoring that environment for 
the absorption of a photon. Once a photon has been detected the 
coupling to the environment is switched off. If the coupling is 
maintained until a photon has been absorbed, then this procedure removes 
exactly one photon from the field.

The paper is organized as follows. Section~\ref{sec:linabs} briefly 
reviews the main properties of linear absorption. The 
master equation describing the evolution of the field in the absorber 
is introduced. These results are used in Section~\ref{sec:adapabs}
where we give the evolution of the field in adaptive absorption, and in 
Section~\ref{sec:implem} where we give a possible implementation. 
In Section~\ref{sec:prop}, we study the evolution under adaptive 
absorption for some relevant input states. Section~\ref{sec:meas}
treats the question obtaining information about the input
field from the time at which the absorption occurs. Finally, in 
Section~\ref{sec:concl} we present some further elaborations and conclude 
the paper.

\section{Linear absorption}\label{sec:linabs}

Absorption in a linear medium is characterized by a reduction of
optical field amplitude by a factor $\eta^{\frac{1}{2}}$ which is 
independent of the field amplitude. A coherent state $\ket{\alpha}$ after being 
partially absorbed during its propagation through, for example, an 
optical fiber is transformed to $\ket{\eta^{\frac{1}{2}}\alpha}$. This is formally 
the same relation as between the input and one of the output arms of a 
beam splitter. The P-representation (see Eq.~\eqref{eq:p-dis} below) 
allows us to extend the previous formal 
analogy to all input states. 
The beam-splitter also serves as a model for  phase-sensitive 
damping \cite{leonh93,kim95} by feeding the 
second input port with a squeezed state.

The action of beam splitters is widely studied (see for 
example \cite{campos89}). Some properties which are relevant in our 
context are: a) The output modes are in general 
entangled and therefore the transmitted state is a mixed state.
b) Coherent states are an exception to the previous statement.
c) When the vacuum and $n$ photons are fed into the input ports,
the  photon number distribution in the output ports follows 
the same statistical law for distributing $n$ classical particles: i.e. 
from $n$ photons the absorber will let through
$m$ photons at random with probability $\eta$ for each, while absorbing $(m-n)$
with probability $1-\eta$. The resulting probability distribution for the 
transmitted field is,
\begin{equation}
    p_{BS}(m)=\left(
    \begin{array}{c}
        n  \\
        m
    \end{array}
    \right) \eta^m (1-\eta)^{(n-m)}.
    \label{eq:prdisnp}
\end{equation}
d) Vacuum fluctuations from the input port are added to the 
 the transmitted field. This illustrates the fact that dissipation 
 is always accompanied by extra fluctuations.

The beam splitter  analogy gives us a relation between the input 
and output photonic states. While the field is in contact with 
the absorbing medium, its 
evolution is non-unitary because of the losses, but 
may be described by the master equation
\begin{equation}
 \frac{d \rho}{dt}=\Gamma (2 a \rho (t) a^\dagger- a^\dagger a \rho (t)-
 \rho (t)a^\dagger a)\mbox{,}
    \label{eq:mastereq}
\end{equation}
where $a$ is the annihilation operator of the field-mode. 
This is the master equation of a harmonic oscillator coupled to a 
broad-bandwidth bath \cite{barnett97}.
It is useful to separate the evolution superoperator into two terms 
defined by
\begin{eqnarray}
    \hat{J} \rho& = & a \rho a^\dagger
    \label{eq:J}  \\
    \hat{L} \rho& = & -\frac{1}{2}[a^\dagger a\rho+\rho a^\dagger a]
    \label{eq:L}
\end{eqnarray}
 The master equation \eqref{eq:mastereq} 
 can now be rewritten in the compact form,
\begin{equation}
    \dot{\rho}=2 \Gamma (\hat{J}+\hat{L})\rho\mbox{.}
    \label{eq:meqcompact}
\end{equation}
A solution of this differential equation may be expressed in a 
number of ways including  \cite{carmich93}, 
\begin{eqnarray}
    \rho(t) & = & \sum_{m=0}^{\infty}\int_{0}^{t} 
    \dt_{m}\int_{0}^{t_{m}} \dt_{m-1}\cdots \int_{0}^{t_{2}} \dt_{1}
    \mbox{e}^{2\Gamma \hat{L}(t-t_{m})} \nonumber \\
     & \times &  2\Gamma\hat{J}\mbox{e}^{2\Gamma \hat{L}(t_{m}-t_{m-1})} \cdots
     2\Gamma\hat{J}\mbox{e}^{2\Gamma \hat{L} t_{1}} \rho(0)
    \label{eq:fsol}  
\end{eqnarray}

This formal solution  allows for an
interpretation in terms of 
conditioned evolution in agreement with quantum measurement theory.
The solution is formed from terms of the form,
\begin{eqnarray}
    \bar{\rho_{c}}(t)&=&\exp[2\Gamma \hat{L}(t-t_{N})] 
    2\Gamma\hat{J}\exp[2\Gamma \hat{L}(t_{N}-t_{N-1})]
     \cdots \nonumber \\
   & &  \cdots 2\Gamma\hat{J}\exp[2\Gamma \hat{L} t_{1}]\rho(0)\mbox{.}
     \label{eq:rhocunnorm}
\end{eqnarray}
Each occurrence of $\hat{J}$ corresponds to 
the loss of a photon. The exponential terms account for the 
non-unitary evolution of the system between photon absorptions.
It follows that
\begin{equation}
    \rho_{c}(t)=\frac{\bar{\rho_{c}}(t)}{\Tr{\bar{\rho_{c}}(t)}}
    \label{eq:rhoc}
\end{equation}
can be interpreted as the  evolved density matrix 
conditioned by the loss of $N$ photons to 
the environment at times $t_{1},t_{2},\ldots, t_{N}$.
The norm of $\bar{\rho_{c}}(t)$ gives 
the probability of this particular sequence of detections to occur. 
The unconditional evolution described in Eq.~\eqref{eq:fsol} is then 
obtained by averaging over all possible histories, i.e. 
summing over the number of lost photons $m$,
and integrating over their corresponding absorption times $t_{1}\ldots t_{m}$.
The action of the superoperator $\hat{J}$ is sometimes referred to as a 
`quantum jump', since it produces a finite change in the state in an 
infinitesimal time. In contrast, when no quantum jump occurs, the change
is  infinitesimal but not unitary.

A coherent state after the loss of a photon and 
 after a time $\Delta t$ without losses is transformed, respectively, 
 as,
\begin{eqnarray}
   \hat{J} \ketbrad{\alpha}& = & |\alpha|^2 \ket{\alpha}\bra{\alpha}
    \label{eq:jalpha} \mbox{,} \\
  \mbox{e}^{2\Gamma \hat{L} \Delta t} \ketbrad{\alpha} & = & 
  \mbox{e}^{-|\alpha|^2(1-\exp(-2\Gamma \Delta t))} \nonumber \\
 & &\times \ketbrad{\alpha \mbox{e}^{-\Gamma \Delta t}}
    \label{eq:exLalpha}
\end{eqnarray}

One immediately notices that for an initial coherent state 
$\rho(0)=\ketbrad{\alpha}$, each integrand in 
Eq.~\eqref{eq:fsol}
is proportional to $\ketbrad{\alpha 
  \mbox{e}^{-\Gamma  t}}$. Hence, the 
  unconditioned state at a time $t$ is also $\ket{\alpha 
  \mbox{e}^{-\Gamma  t}}$.
This shows the equivalence of the  master equation ~\eqref{eq:mastereq}
to the beam splitter from above (with $\eta=\mbox{e}^{-2\Gamma t}$.)
The master equation description and its interpretation in terms of 
conditioned evolution
gives us a tool to formalize adaptive absorption in the following section.
 
\section{Adaptive absorption}\label{sec:adapabs}
In this section we show how to remove a single photon from an 
field-mode prepared in any state. For this we need to couple our field
to the broad-bandwidth environment as in the previous section.
The state will then evolve according to 
Eq.~\eqref{eq:fsol}, or Eq.~\eqref{eq:rhocunnorm} if we are able to 
keep track of the jump times. Now suppose that we switch 
off the coupling to the environment immediately after the a single 
photon has been absorbed (see below for a possible implementation). 
In this case the conditional state after a photodetection at time 
$t_{1}<t$ will be
\begin{equation}
    \bar{\rho_{c}}(t)=2\Gamma\hat{J}\mbox{e}^{2\Gamma \hat{L}t_{1}}\rho(0)\mbox{.}
    \label{eq:rhocond}
\end{equation}
We can predict the evolution of this system in which a photon may be 
absorbed at any time before $t$, or no photon is absorbed. The 
corresponding unconditioned normalized density matrix is,
\begin{equation}
   \rho(t)=
   \exp[2\Gamma \hat{L} t]\rho(0)+\int_{0}^{t}\dt_{1}2\Gamma\hat{J}
   \exp[2\Gamma \hat{L}t_{1}]\rho(0)\mbox{.}
   \label{eq:rhoc1ph}
\end{equation}

It is worth noting that this evolution is explicitly non-Markovian.
The future evolution depends not only on the state but also on its 
previous history. This can also be seen by differentiating 
the previous equation,
\begin{eqnarray}
    \dot{\rho}(t)&=&2 \Gamma (\hat{J}+\hat{L})\mbox{e}^{2\Gamma 
    \hat{L} t}\rho(0) \nonumber \\
    &=& 2 \Gamma (\hat{J}+\hat{L}) \bar{\rho}_{\mbox{no 
    detection}}(t),
    \label{eq:meqdiff}
\end{eqnarray}
where the `no detection subscript denotes the density operator for 
which no photon has been absorbed.

Before examining the properties of Eq.~\eqref{eq:rhoc1ph}, we  
describe a scheme to implement the adaptive absorption.

\section{Implementation of Adaptive absorption}\label{sec:implem}
\begin{figure}%[!hbp]
    \centerline{\psfig{figure=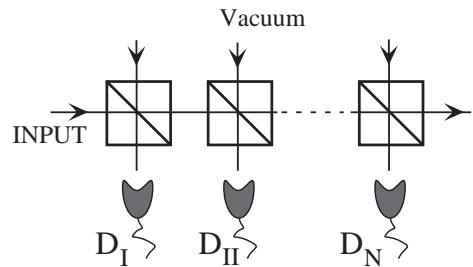,width=2.4in}}
    \caption{Beam splitter model for a field-mode coupled to an
     environment}
    \label{fig:id1}
\end{figure}
The basic idea is depicted in 
Fig.~\ref{fig:id1}.
The input signal is send through a series of weak beam 
splitters.  Their reflection coefficient is very small, so that there is
a  negligible probability for more than a single photon to be reflected.
A photodetector is placed in the weakly coupled arm of each beam 
splitter. What we have described until now, is our field 
mode coupled to the broad-bandwidth environment. The corresponding 
evolution is then described by Eq.~\eqref{eq:fsol} or 
Eq.~\eqref{eq:rhocunnorm} depending on whether the firing times of the 
detectors are known or not. In the latter case 
the input-output relation is equivalent to a single beam 
splitter of finite reflectivity. In particular this means that all 
vacuum contributions interfere to give a single vacuum contribution, 
so that the act of continuously monitoring does not entail 
extra fluctuations.

We seek to enforce the conditional evolution.
This means that if a signal is registered in the 
first detector $D_{I}$, no further beam splitters should be interposed.
On the other hand, if the first detector does not fire, the light is 
sent to the second beam splitter and the photodetector $D_{II}$
is checked for a firing signal. The process is repeated
until a photodetector fires.

\begin{figure}[!hbp]
    \centerline{\psfig{figure=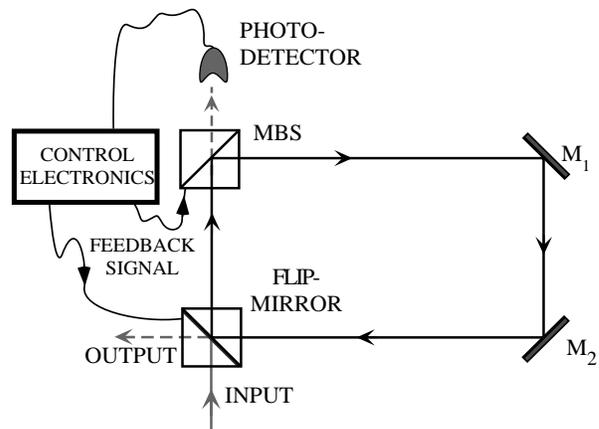,width=3.2in}}
    \caption{Possible experimental setup for single photon extraction by 
    adaptive absorption.}
    \protect\label{fig:impl1}
\end{figure}

In order to implement this procedure, the experimental 
arrangement in Fig.~\ref{fig:impl1} might be used. 
In this scheme only four optical elements  and a single 
photodetector are needed.
When a light pulse is trapped in
the loop enclosed between the modulated beam splitter $MBS$, the 
mirrors $M_{1}$ and $M_{2}$, and the flip mirror, it is effectively 
coupled to the environment through the modulated beam splitter $MBS$. 
The value of its reflectivity is set close to one and  
can be slightly increased via an electric current to make it fully 
reflective.
In order to effectively feed the initial state 
into the loop the flip mirror has to open. 

The photodetector keeps 
track of the photon losses and as soon as it fires, it sends a feedback 
signal to the flip-mirror so that the pulse can exit the 
loop without any further losses. 
Alternatively, the coupling
of the field to the environment can be switched off 
by turning the $MBS$ fully reflective.
The flip mirror is then switched whenever the output is desired.
This last alternative allows the system to provide the unconditioned 
evolved states. Otherwise different sub-ensembles will emerge at the 
output at different times.  
Depending on the response times of the 
photodetector, control electronics, modulated beam-splitter,
and the flip-mirror one can add a 
delay line within the loop, 
for example between the two mirrors $M_{1}$ and $M_{2}$. 

Any real experiment  will inevitably be only an approximation to 
the model described above. In practice every beam splitter has a finite 
reflectivity and has some internal losses, and detectors have 
efficiencies below one. Inclusion of such effects will degrade the 
performance of the device.

\section{Properties of the evolved state}\label{sec:prop}

In this section we list some properties of the evolved state.
Eqs.~\eqref{eq:rhocond} and \eqref{eq:rhoc1ph} give the 
conditional and unconditional evolution for an arbitrary initial 
state and we now present the evolution for two important examples.

\subsection{Coherent state}\label{sec:coherent.}

We consider first the important example of an initial coherent state 
$\rho^{\alpha}(0)=\ketbrad{\alpha}$.  
As shown in Eq. (\ref{eq:jalpha}), a coherent
state remains unaffected under the action of the jump 
superoperator $\hat{J}$. This can be understood in the context of
the implementation in Fig.~\ref{fig:impl1}. As already mentioned,
coherent states have the peculiarity that when sent 
through a beam splitter the two outcoming beams 
are not entangled. This implies that the transmitted beam is the same
independently of the measurement outcome 
at the photodetector.
In any case, while coupled to the environment,
the coherent amplitude is damped following 
Eq.~\eqref{eq:exLalpha}.
Thus the output field is determined only by the time 
which the states spend in contact with the environment (inside the 
loop in Fig.~\ref{fig:impl1}), that is the time until the 
first jump occurs. The conditioned state at the output when the 
detector fires at $t_{1}$ is given by
\begin{equation}
    \rho_{c}^{\alpha}(t)=
    \ketbra{\alpha\mbox{e}^{-\Gamma  t_{1}}}{\alpha \mbox{e}^{-\Gamma t_{1}}}
    \mbox{.}
    \label{eq:rhoct1}
\end{equation}
This occurs with the probability
\begin{equation}
 p(t_{1}|\alpha) \dt =2 \Gamma |\alpha|^{2}\mbox{e}^{-2\Gamma t_{1}}
 \exp[-|\alpha|^2(1-\mbox{e}^{-2\Gamma  t_{1}})] \dt\mbox{.}
    \label{eq:prbonephcoh}
\end{equation}

Following Eq.~\eqref{eq:rhoc1ph} the unconditional evolution of the 
coherent state is
\begin{eqnarray}
    \rho^{\alpha}(t)& = &\exp[-|\alpha|^2(1-\mbox{e}^{-2\Gamma  t})]
    \ketbrad{\alpha\mbox{e}^{-\Gamma  t}}
    \nonumber \\
   &+&\int_{0}^{t}\dt_{1}2 \Gamma |\alpha|^{2}\mbox{e}^{-2\Gamma 
   t_{1}} \nonumber \\
   & &\times\exp[-|\alpha|^2(1-\mbox{e}^{-2\Gamma  t_{1}})]
    \ketbrad{\alpha\mbox{e}^{-\Gamma  t_{1}}}\mbox{.}
    \label{eq:rhcohuncon}
\end{eqnarray}
The P-representation of this state, defined by the equation
\begin{equation}
    \rho^{\alpha}(t)=\int \mbox{d}^{2}\beta P^{\alpha}(\beta,t)\ketbrad{\beta}
    \mbox{,}
    \label{eq:p-dis}
\end{equation}
is given by
\begin{eqnarray}
    && P^\alpha(\beta,t)=\left\{
    \exp[-|\alpha|^2(1-\mbox{e}^{-2\Gamma  t})]|\beta|^{-1}
     \right.
    \nonumber \\
   &\times&\left.\delta(|\beta|-|\alpha|\mbox{e}^{-\Gamma t})
   +2\mbox{e}^{(|\beta|^{2}-|\alpha|^2)}
   H(|\beta|-|\alpha|\mbox{e}^{-\Gamma t})\right.\nonumber \\
    &\times& \left. H(|\alpha|-|\beta|)\right\} \delta(\arg\alpha-\arg\beta)\mbox{,}
    \label{eq:pbeta}
\end{eqnarray}
where $H(x)$ is the Heaviside step function.
The existence and form of this P-function allow us to draw some 
conclusions: 1) The state produced in this way will have essentially 
classical properties. 2) The removal of the photon does not change 
the argument of the coherent state in that the P-function is zero for 
phases other than $\arg\alpha$.

\begin{figure}%[!hbp]
    \centerline{\psfig{figure=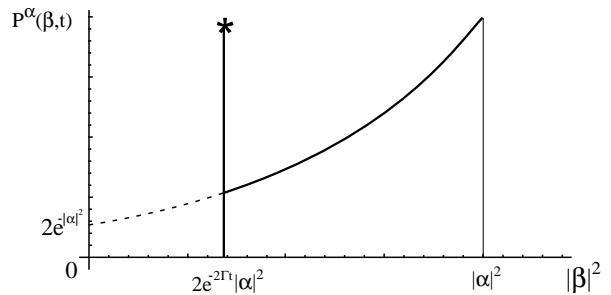,width=3.4in}}
    \caption{P-distribution for a coherent state 
    $\ket{\alpha}$ evolved until time $t$ (full line). 
    $P^\alpha(\beta,t)$ is non-vanishing
    only at the fixed phase $\arg \beta=\arg\alpha$.}
    \protect\label{fig:pdis}
\end{figure}

 The plot of the P-function $P^\alpha(\beta,t)$ in  Fig.~\ref{fig:pdis}
gives us an intuitive picture of the evolution of the unconditioned state.
The initial state $\ket{\alpha}$  transforms after a time $t$ into a 
mixture of coherent states with damped amplitudes $|\beta|\in 
[|\alpha|\mbox{e}^{-\Gamma t},|\alpha|)$. The contribution in the 
mixture of each coherent state also decays exponentially.
The singular peak ($*$ in Fig.~\ref{fig:pdis}) corresponds to the 
case where the detector has not fired, and it takes the required 
weight to keep $P^\alpha(\beta,t)$ normalized.

\subsection{Number states and photon statistics}\label{sec:number}
Another important basis-state is the number state $\ket{n}$.
With probability
\begin{equation}
 p(t_{1}|n) \dt =2\Gamma n \mbox{e}^{-2\Gamma n t_{1}}\dt
    \label{eq:prbonenumber}
\end{equation}
the number state $\ket{n}$ will lose one photon at time $t_{1}$,
to produce the conditioned state 
\begin{equation}
\rho_{c}^n(t)=\ketbrad{n-1}
    \label{eq:rhocnum}
\end{equation}
for times $t>t_{1}$.
This state is independent of the 
time $t_{1}$ in which the jump occurs. As a result, the unconditioned 
state when no information on the detection events is available, 
reduces to the simple form,
\begin{equation}
    \rho^{n}(t) = \mbox{e}^{-2 n \Gamma t}
    \ketbrad{n}+ (1-\mbox{e}^{-2 n \Gamma t}) \ketbrad{n-1}\mbox{.}
    \label{eq:rhcnumcon}
\end{equation}

In contrast to the coherent state case, the number state soon evolves 
into a pure state.
The probability of not extracting one photon from $\ket{n}$ falls 
exponentially with time, thus the extraction 
is bound to occur given enough time. After long times $t\rightarrow\infty$ a single photon is 
removed from the signal deterministically (unless, of course, there 
was no photon to begin with).

This last remarks are in fact valid for any initial state
$\rho(0)=\sum_{n,n'=0}^{\infty}\rho_{n,n'}\ketbra{n}{n'}$. 
Indeed, the photon number probability distribution at a time $t$ is given by
\begin{eqnarray}
    p_{n}(t)  &=&  \bra{n}\rho(t)\ket{n} \nonumber \\ 
    &=&\mbox{e}^{-2 n \Gamma t} 
    p_{n}(0)+(1-\mbox{e}^{-2 n \Gamma t})p_{n+1}(0).
    \label{eq:phstat}
\end{eqnarray}
And for $t\rightarrow\infty$ we have
\begin{equation}
    p_{n}(\infty)  = p_{n+1}(0)+p_{0}(0) \delta_{n,0}
    \label{eq:phstatin}
\end{equation}
which means that the probability distribution is shifted one step to 
smaller values of $n$. The initial vacuum component is of course not 
shifted and still receives the contribution from the initial 
one-photon component.  In this limit of large times the mean and 
variance of the photon number can be easily calculated using 
Eq.~\eqref{eq:phstatin},
\begin{eqnarray}
    \bar{n}(\infty) & = &\bar{n}(0) +p_{0}(0)-1 
   \mbox{,}
    \label{eq:nbar}  \\
    \Delta n^2(\infty) & = & \Delta 
    n^2(0)-p_{0}(0)[2\bar{n}(0)+p_{0}(0)-1]  \nonumber \\
    &=&\Delta n^2(0)-p_{0}(0)[\bar{n}(0)+\bar{n}(\infty)]\mbox{.}
    \label{eq:varn}
\end{eqnarray}

The mean photon number decreases by an amount equal to the initial
probability of having a non-vacuum signal. The variance
decreases by an amount dictated by the vacuum probability and also by
the mean photon number.
From the results in the previous section we know that adaptive 
absorption cannot produce non-classical light from initially 
classical light, in the sense that it preserves the positivity of the 
P-function.  Despite this, it still enables us to change substantially
the photon statistics of the incoming field. The normally ordered 
variance of the number operator is
\begin{eqnarray}
    :\!\Delta n^2\!:(\infty)&=&\Delta n^2(\infty)-\bar{n}(\infty)
    \nonumber \\
   &=& :\!\Delta n^2\!:(0)+1 -p_{0}(0)[2\bar{n}(0)+p_{0}(0)]\mbox{.}
    \label{eq:novar}
\end{eqnarray}
It is now easy to find an example in which an 
initial super-Poissonian distribution ($:\!\Delta n^2\!:>0$) 
gives a sub-Poissonian output field.
An incoming field $\rho^A(0)$ with 
$p^A_{n}(0)=p_{0}\delta_{n,0}+(1-p_{0})\delta_{n,N}$ has a normal 
ordered variance $:\!\Delta n^2\!:= N(1-p_{0})(N p_{0}-1)$ which is 
positive for $N>p_{0}^{-1}$.  After long times the corresponding 
output field will have the normal ordered variance,
\begin{equation}
     :\!\Delta n^2\!:^{A}(\infty)=(N-1)(1-p_{0})[(N-1)p_{0}-1]\mbox{,}
    \label{eq:exnov}
\end{equation}
which becomes negative for $1<N<p_{0}^{-1}+1$. Summarizing, for every 
non-integer value of $p_{0}^{-1}$ there exists an integer $N$
which fulfills the inequalities $p_{0}^{-1}<N<p_{0}^{-1}+1$ 
so that the corresponding super-Poissonian input
 field becomes sub-Poissonian after the adaptive absorption procedure.

\section{Adaptive absorption as photon number 
measurement}\label{sec:meas}

Along the course of this paper we have investigated adaptive absorption as a 
way to manipulate photonic states. By now we have expressed in many 
ways the idea that the induced evolution 
is non-unitary and that the coupling to the environment plays a 
crucial role in this manipulation. In quantum 
information theory this type of evolution is always 
accompanied by an outflow of information to the 
environment. In this section we show that indeed adaptive 
absorption can also be understood as a procedure to measure the input
photonic state. 
The balance between information gain and 
disturbance is highly non-trivial and delicate. Moreover, when the 
system under study is an optical field-mode we find practical limitations 
in that many of the interesting quantum operations require non-linear 
optical devices. Here we present a potentially valuable tool which
only makes use of linear 
optical elements. 

In Section~\ref{sec:number} we have seen that for long times, the 
effect of adaptive absorption  on the photon number 
distribution was to shift it down by one. Now we will see that 
 by keeping track of the detection times $t_{1}$ we can obtain 
information on the initial photon-number. Taking into account that the 
probability of removing one photon from a state of $n$ photons is 
the same as for the analogous classical  problem, it is easy to 
understand that the larger the initial photon number, the easier 
(i.e., the faster) it is to extract a single photon. 
The time $t_{1}$ at which the photon is absorbed is our measurement outcome. 
In fact, at each time step we are effectively performing the 
measurement defined by the
\emph{positive operator-valued measure} or POVM
\cite{peres93a}  $\{\Pi_{1}= 2 \dt\Gamma 
a^{\dagger}a,\Pi_{0}=\openone -\Pi_{1}\}$. Since the absorption POVM 
$\Pi_{1}$ is infinitesimally small, most of the times we get the 
``0'' outcome. The value of $t_{1}$ conveys  the 
 result of the measurements up to $t_{1}$.
In order to know the information obtained from the knowledge of $t_{1}$,
the central quantity is the  probability $p(n|t_{a})$ of 
having a certain number $n$ of photons in the input given that the 
measurement outcome is $t_{1}=t_{a}$. By 
invoking Bayes theorem we can  write this conditional 
probability as
\begin{equation}
p(n|t_{a})=\frac{p(t_{a}|n) p_{n}}{p(t_{a})}=
\frac{p(t_{a}|n)}{\sum_{m=1}^{\infty}p(t_{a}|n)}\mbox{,}
    \label{eq:bayes}
\end{equation}
where $p(t_{a})=\sum_{n=0}^{\infty} p_{n}p(t_{a}|n)$ 
and we assume for definiteness a flat \emph{a priori} probability distribution 
$p_{n}=p$  $\forall n$.
Use of the conditional probability given in Eq.~\eqref{eq:prbonenumber} 
gives
\begin{equation}
p(n|t_{a})= n \mbox{e}^{-2\Gamma (n+1) t_{a}} (\mbox{e}^{2\Gamma 
t_{a}}-1)^{2}\mbox{.}
    \label{eq:bayes2}
\end{equation}

\begin{figure}%[!hbp]
    \centerline{\psfig{figure=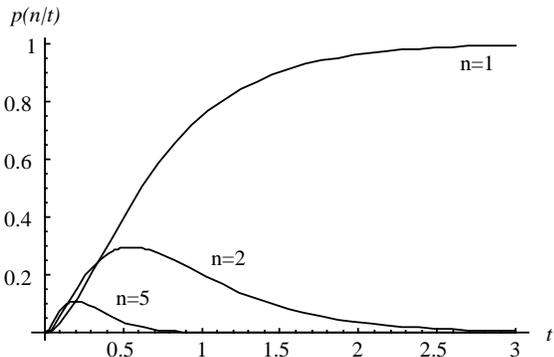,width=3in}}
    \caption{$p(n|t_{a})$ for $\Gamma=1$ and three different values of
    $n=\{1,2,5\}$.}
    \protect\label{fig:pnt}
\end{figure}

We note that summation of this expression over $n$ from $1$ to 
$\infty$ gives unity as it should. The exponential dependence on $n$ 
may provide the means to discriminate between states with widely 
differing photon numbers without detecting all of the photons 
(see Figure~\ref{fig:pnt}). The technique cannot, of course, determine 
the number of photons without error. The benefit of this technique as 
a measurement, however, is likely to be in those situations where we 
can only afford to weakly perturb the mode. A possible application of 
this idea to quantum cryptography is discussed briefly in the 
concluding section.

\section{Conclusions}\label{sec:concl}

In this paper we have presented the idea of extracting a single 
photon from an arbitrary single-mode field via adaptive absorption.
The extraction of a single photon can also be achieved by post-selection,
that is by setting up a beam splitter and considering at the end only 
those situations in which a single photon was removed. Of course this 
is not deterministic and would in fact lead to the possibility of 
removing more than one photon. The fact that one can stop the absorption process as soon 
as one photon is absorbed is the novel aspect of this work. For example 
this possibility  has given rise to a novel eavesdropping attack
on quantum key distribution (QKD) \cite{calsam01}. 
There are two factors that might render insecure many of the
current implementations of QKD \cite{hut95}:
the signals are weak coherent pulses instead of single photons 
and the communication channels have very high losses. 
The eavesdropper  can acquire the secret key whenever he
succeeds in extracting a single photon from the multiphoton part of the signal.
Adaptive absorption allows the eavesdropper to take advantage of this 
security gap by using current technology.

By repeatedly applying the single-photon extraction one can extract 
any number of photons in a controlled way. Moreover, as we have seen in the 
previous section, in process of extracting single-photons  
we also obtain information on the input system. Thus adaptive 
absorption implements weak measurements, where the measured system is 
only slightly disturbed.

Adaptive absorption is a simple example of a general idea that can 
give rise to many other applications. Another example, that falls 
outside the scope of this paper, is adaptive amplification as a means to
add a single excitation into a mode. 
This general idea is to use the results of a continuous 
measurement to modify the system Hamiltonian or its coupling to the
measurement apparatus. This line of action may lead to novel quantum 
operations or ease the realization of operations
with too high technological demands.

\section{Acknowledgements}\label{ackw}
S. M. B.  thanks the Royal Society of Edinburgh and the Scottish Executive 
Education and Lifelong Learning Department for financial support.
J.C. and K-A.S. acknowledge the Academy of Finland (project 4336)
and the European Union IST EQUIP Programme for financial support.

%\begin{multicols}{1}

%\bibliography{/root/tex/STYLES_ETC/norbert.bib}
%\bibliographystyle{/root/tex/STYLES_ETC/prsty}

\end{multicols}

\end{document}